\documentclass[12pt]{article}

\usepackage{graphics,psfrag}
\usepackage{amsthm,amssymb,epsfig,amsmath,euscript,array,cite}

\newcounter{multieqs}

\newcommand{\be}{\begin{equation}}
\newcommand{\ee}{\end{equation}}
\newcommand{\eq}[1]{(\ref{#1})}

\newcommand{\bm}[1]{\mbox{\boldmath $#1$}}
\newcommand{\rf}[1]{(\ref{#1})}

\def\bd{\begin{document}}
\def\ed{\end{document}}
\def\nn{\nonumber}
\def\bea{\begin{eqnarray}}
\def\eea{\end{eqnarray}}
\let\bm=\bibitem
\let\la=\label

\def\npb#1#2#3{Nucl. Phys. {\bf{B#1}} #3 (#2)}
\def\plb#1#2#3{Phys. Lett. {\bf{#1B}} #3 (#2)}
\def\prl#1#2#3{Phys. Rev. Lett. {\bf{#1}} #3 (#2)}
\def\prd#1#2#3{Phys. Rev. {D \bf{#1}} #3 (#2)}
\def\cmp#1#2#3{Comm. Math. Phys. {\bf{#1}} #3 (#2)}
\def\cqg#1#2#3{Class. Quantum Grav. {\bf{#1}} #3 (#2)}
\def\nppsa#1#2#3{Nucl. Phys. B (Proc. Suppl.) {\bf{#1A}}#3 (#2)}
\def\ap#1#2#3{Ann. of Phys. {\bf{#1}} #3 (#2)}
\def\ijmp#1#2#3{Int. J. Mod. Phys. {\bf{A#1}} #3 (#2)}
\def\rmp#1#2#3{Rev. Mod. Phys. {\bf{#1}} #3 (#2)}
\def\mpla#1#2#3{Mod. Phys. Lett. {\bf A#1} #3 (#2)}
\def\jhep#1#2#3{J. High Energy Phys. {\bf #1} #3 (#2)}
\def\atmp#1#2#3{Adv. Theor. Math. Phys. {\bf #1} #3 (#2)}

\def\N{{\cal N}}
\def\sst{\scriptscriptstyle}
\def\thetabar{\bar\theta}
\def\Tr{{\rm Tr}}
\def\one{\mbox{1 \kern-.59em {\rm l}}}

\def\a{\alpha}       
\def\b{\beta}         
\def\g{\gamma}  \def\G{\Gamma}  \def\dc{{\dot\gamma}}  
\def\d{\delta}  \def\D{\Delta}  \def\ddt{\dot\delta}  
\def\e{\epsilon}        \def\vare{\varepsilon}  
\def\f{\phi}    \def\F{\Phi}    \def\vvf{\f}  
\def\h{\eta}  
\def\k{\kappa}  
\def\l{\lambda} \def\L{\Lambda}  
\def\m{\mu} \def\n{\nu}  
\def\o{\omega}  
\def\p{\pi} \def\P{\Pi}  
\def\r{\rho}  
\def\s{\sigma}  \def\S{\Sigma}  
\def\t{\tau}  
\def\th{\theta} \def\Th{\Theta} \def\vth{\vartheta}  
\def\X{\Xeta}  
\def\z{\zeta}  

\def\na{\nabla}  

\def\cA{{\cal A}} \def\cB{{\cal B}} \def\cC{{\cal C}}  
\def\cD{{\cal D}} \def\cE{{\cal E}} \def\cF{{\cal F}}  
\def\cG{{\cal G}} \def\cH{{\cal H}} \def\cI{{\cal I}}  
\def\cJ{{\cal J}} \def\cK{{\cal K}} \def\cL{{\cal L}}  
\def\cM{{\cal M}} \def\cN{{\cal N}} \def\cO{{\cal O}}  
\def\cP{{\cal P}} \def\cQ{{\cal Q}} \def\cR{{\cal R}}  
\def\cS{{\cal S}} \def\cT{{\cal T}} \def\cU{{\cal U}}  
\def\cV{{\cal V}} \def\cW{{\cal W}} \def\cX{{\cal X}}  
\def\cY{{\cal Y}} \def\cZ{{\cal Z}}

\def\ua{\underline{\alpha}}  
\def\ub{\underline{\phantom{\alpha}}\!\!\!\beta}  
\def\uc{\underline{\phantom{\alpha}}\!\!\!\gamma}  
\def\um{\underline{\mu}}  
\def\ud{\underline\delta}  
\def\ue{\underline\epsilon}  
\def\una{\underline a}\def\unA{\underline A}  
\def\unb{\underline b}\def\unB{\underline B}  
\def\unc{\underline c}\def\unC{\underline C}  
\def\und{\underline d}\def\unD{\underline D}  
\def\une{\underline e}\def\unE{\underline E}  
\def\unf{\underline{\phantom{e}}\!\!\!\! f}\def\unF{\underline F}  
\def\unm{\underline m}\def\unM{\underline M}  
\def\unn{\underline n}\def\unN{\underline N}  
\def\unp{\underline{\phantom{a}}\!\!\! p}\def\unP{\underline P}  
\def\unq{\underline{\phantom{a}}\!\!\! q}  
\def\unQ{\underline{\phantom{A}}\!\!\!\! Q}  
\def\unH{\underline{H}}  
  
\def\As {{A \hspace{-6.4pt} \slash}\;}  
\def\bs {{b \hspace{-6.4pt} \slash}\;}  
\def\Ds {{D \hspace{-6.4pt} \slash}\;}  
\def\ds {{\del \hspace{-6.4pt} \slash}\;}  
\def\ss {{\s \hspace{-6.4pt} \slash}\;}  
\def\ks {{ k \hspace{-6.4pt} \slash}\;}  
\def\ps {{p \hspace{-6.4pt} \slash}\;}   
\def\xs {{x \hspace{-6.4pt} \slash}\;}  
\def\pas {{{p_1} \hspace{-6.4pt} \slash}\;}  
\def\pbs {{{p_2} \hspace{-6.4pt} \slash}\;}  
  
\def\Dh{\hat{D}}
\def\Gh{\hat{G}}
\def\Fh{\hat{F}}
\def\Ph{\hat{P}}
\def\Rh{\hat{R}}
\def\Vh{\hat{V}}  
\def\Xh{\hat{X}} 
 
\def\ah{\hat{a}}
\def\gh{\hat{g}} 
\def\hh{\hat{h}}
\def\uh{\hat{u}}  
\def\xh{\hat{x}}  
\def\yh{\hat{y}}  
\def\ph{\hat{p}}  
\def\xih{\hat{\xi}}  
\def\chih{\hat{\chi}}

\def\psit{\tilde{\psi}}  
\def\Psit{\tilde{\Psi}}   
\def\Psibt{\tilde{\bar{Psi}}}  

\def\Phit{\tilde{\Phi}}   
\def\Phitb{\overline{\tilde{Phi}}}  
\def\tht{\tilde{\th}}  
\def\lt{\tilde{\l}}
\def\chit{\tilde{\chi}}   

\def\At{\tilde{A}}
\def\Dt{\tilde{D}}
\def\Ft{\tilde{F}}
\def\Qt{\tilde{Q}}  
\def\Rt{\tilde{R}}  
\def\Mt{\tilde{M }}  
\def\Nt{\tilde{N}}   
\def\Xt{\tilde{X}}
 
\def\at{\tilde{a}}  
\def\htt{\tilde{h}} 
\def\st{\tilde{s}}  
\def\ft{\tilde{f}}
\def\gt{\tilde{g}}
\def\pt{\tilde{p}}  
\def\qt{\tilde{q}}  
\def\vt{\tilde{v}}  
\def\nt{\tilde{n}}  
\def\ut{\tilde{u}} 
\def\xt{\tilde{x}} 
\def\yt{\tilde{y}} 
\def\Psit{\tilde{\Psi}}
\def\vphit{\tilde{\varphi}}  

\def\delb{\bar{\partial}}  
\def\thb{\bar{\theta}}
\def\mub{\bar{\mu}}
\def\lamb{\bar{\l}}
\def\psib{\bar{\psi}}
\def\sb{\bar{\sigma}}
\def\xib{\bar{\xi}}
\def\chib{\bar{\chi}}

\def\Phib{\bar{\Phi}}
\def\Lamb{\bar{\Lambda}}

\def\Ab{{\overline A}} \def\Bb{{\overline B}} \def\Cb{{\overline C}}  
\def\Db{{\overline D}} \def\Eb{{\overline E}} \def\Fb{{\overline F}}  
\def\Gb{{\overline G}} \def\Hb{{\overline H}} \def\Ib{{\overline I}}  
\def\Jb{{\overline J}} \def\Kb{{\overline K}} \def\Lb{{\overline L}}  
\def\Mb{{\overline M}} \def\Nb{{\overline N}} \def\Ob{{\overline O}}  
\def\Pb{{\overline P}} \def\Qb{{\overline Q}} \def\Rb{{\overline R}}  
\def\Sb{{\overline S}} \def\Tb{{\overline T}} \def\Ub{{\overline U}}  
\def\Vb{{\overline V}} \def\Wb{{\overline W}} \def\Xb{{\overline X}}  
\def\Yb{{\overline Y}} \def\Zb{{\overline Z}}  

\def\fb{{\overline f}}
\def\gb{{\overline g}}
\def\mb{{\overline m}}
\def\lb{{\overline l}}
\def\yb{{\overline y}}

\def\bz{{\bar z}}
\def\bA{{\bar A}}
\def\bB{{\bar B}}

\def\ba{{\bf a}} 
\def\bk{{\bf k}}  
\def\bl{{\bf l}}  
\def\bp{{\bf p}}  
\def\bq{{\bf q}}  
\def\br{{\bf r}}
\def\bt{{\bf t}}
\def\bu{{\bf u}}
\def\bv{{\bf v}}
\def\bx{{\bf x}}  
\def\by{{\bf y}}  
\def\bR{{\bf R}}  
\def\bN{{\bf N}}  
\def\bbN{{\bf \bar{N}}}  
\def\bV{{\bf V}}

\def\bone{{\bf 1}}  

\def\va{{\vec a}}
\def\vk{{\vec k}}
\def\vp{{\vec p}}
\def\vq{{\vec q}}
\def\vx{{\vec x}}
\def\vy{{\vec y}}
\def\vu{{\vec u}}
\def\vv{{\vec v}}

\def\vs{{\vec \sigma}}
\def\vtau{{\vec \tau}}

\newcommand{\ov}[1]{\overrightarrow{#1}}
  
\def\da{{\dot a}}
\def\db{{\dot b}}

\def\d{\delta}\def\D{\Delta}\def\ddt{\dot\delta}  
  
\def\pa{\partial} \def\del{\partial}  
\def\xx{\times}  
\def\uno{\mbox{1 \kern-.59em {\rm l}}}    
  
\def\trp{^{\top}}  
\def\inv{^{-1}}  
\def\dag{{^{\dagger}}}  
\def\pr{^{\prime}}  
  
\def\rar{\rightarrow}  
\def\lar{\leftarrow}  
\def\lrar{\leftrightarrow}  
  
\newcommand{\0}{\,\!}      
\def\one{1\!\!1\,\,}  
\def\im{\imath}  
\def\jm{\jmath}  
  
\newcommand{\tr}{\mbox{tr}}  
\newcommand{\slsh}[1]{/ \!\!\!\! #1}  
  
\def\vac{|0\rangle}  
\def\lvac{\langle 0|}  
  
\def\hlf{\frac{1}{2}}  
\def\ove#1{\frac{1}{#1}}  

\def\Box{\square}  
\def\ZZ{\mathbb{Z}}  
\def\RR{\mathbb{R}}   
\def\CC{\mathbb{C}}  
\def\bb#1{{\bf #1}}  
\def\bcomment#1{}  
\def\bfhat#1{{\bf \hat{#1}}}  
\def\VEV#1{\left\langle #1\right\rangle}  

\newcommand{\ex}[1]{{\rm e}^{#1}} \def\ii{{\rm i}}  

\newcommand{\lrbrk}[1]{\left(#1\right)}
\newcommand{\sfrac}[2]{{\textstyle\frac{#1}{#2}}}
 
\def\stw{{\sqrt{2}}}

\def\rf {{\rm f}}
\def\ri {{\rm i}}
\def\rs {{\scriptscriptstyle \rm S}}
\def\rt {{\scriptscriptstyle \rm T}}

\def\rQ {{\scriptscriptstyle \rm \cQ}}
\def\rR {{\scriptscriptstyle \rm \cR}}

\def\cQb{{\cal \Qb}}
\def\cRb{{\cal \Rb}}
\def\cWb{{\cal \Wb}}

\def\fd {{\rm N}}
\def\afd {{\overline{\rm N}}}

\def \II {I\hspace{-.1em}I\hspace{.1em}}
\def \IIA {\mbox{\II A\hspace{.2em}}}
\def \IIB {\mbox{\II B\hspace{.2em}}}
\def \gs {g^s}
\def \ls {\lambda^s}

\def \I {{\cal I}}
\def \qs {q\hspace{-.53em}/\hspace{.15em}}
\def \ks {k\hspace{-.53em}/\hspace{.15em}}
\def \YM {{\mbox{\tiny YM}}}
\def \gym {g_{\YM}}

\def \Lc {\L_c}

\begin{document}

\hfill{DCPT-08/53}

\vspace{20pt}

\begin{center}

{\Large \bf
UV-divergences of Wilson Loops \\
for Gauge/Gravity Duality
}
\vspace{20pt}

{\bf Chong-Sun Chu, Dimitrios Giataganas}

{\em
Centre for Particle Theory and Department of Mathematics,\\
Durham University, South Road,
Durham, DH1 3LE, UK.
}

{\small \sffamily chong-sun.chu@durham.ac.uk,
dimitrios.giataganas@durham.ac.uk }

\vspace{30pt}
{\bf Abstract}
\end{center}

We analyze the structure of the UV divergences of the Wilson loop for a
general gauge/gravity duality.  We find that, due to the
presence of a nontrivial NSNS $B$-field and metric, new
divergences that cannot be subtracted out by
the conventional Legendre transform may arise. We also derive
conditions on the $B$-field and the metric,  which when
satisfied, the leading  UV divergence will
become  linear, and  can be cancelled out by choosing the
boundary condition of the string appropriately.
Our results, together with the recent result of arXiv:0807.5127, where the
effect of a nontrivial dilaton on the structure of UV divergences in
Wilson loop is analysed,
allow us to conclude that Legendre transform is at best capable of
cancelling the linear UV divergences arising from the area of the
worldsheet, but is incapable to handle the divergences
associated with the dilaton or the $B$-field in general. We also solve the
conditions for the cancellation of the  leading linear divergences
generally and  find that many  well-known supergravity backgrounds are
of these kinds, including examples such as
the Sakai-Sugimoto QCD model or $\cN=1$ duality with Sasaki-Einstein
spaces. We also point out that Wilson loop in the
Klebanov-Strassler background have a
divergence associated with the $B$-field
which cannot be cancelled away with the
Legendre transform. Finally we end with some comments on the form of the
Wilson loop operator in the ABJM superconformal Chern-Simons theory.

\setcounter{page}0
\newpage

\section{Introduction}

The AdS/CFT correspondence states the equivalence of string theory on
$AdS_5 \times S^5$ to the $\cN=4$ supersymmetric Yang-Mills
\cite{adscft1,adscft2,adscft3,adscft4}. According to this
correspondence, there exists a map between gauge invariant operators in
the field theory and states in the string theory. The correspondence is
well understood for the case of half BPS local operators where the dual
string states are D-branes in the bulk \cite{cjr,berenstein}. Wilson
loop operator is another class of gauge invariant operator. In the limit
of $N\to \infty$ and large $\l =g^2 N \gg 1$, the expectation value of
a special class
of Wilson loops in the $\cN=4$ SYM theory can be computed using the
supergravity dual picture in terms of a dual
string worldsheet \cite{Mal,rey},\cite{dgo}.
These  Wilson loop operator takes the
form \cite{Mal}
\be \label{wilsono}
W[C] =\frac{1}{N}\Tr \,P\,
\exp\left(\oint_C d\t (iA_\mu \dot x^\mu+\varphi_i\dot y^i)\right),
\ee
where the trace is over the fundamental representation of
the gauge group $G$,
$A_\mu$ are the gauge fields and $\varphi_i$ are the six real
scalars.
The loop $C$ is parametrized by the variables   $(x^\mu(\t), y^i(\t))$, where
$(x^\mu(\t))$ determines the actual loop in four dimensions, and
$(y^i(\t))$ parametrizes the coupling to the scalars. Moreover the
condition
\be\label{basicconstr}
\dot{x}^2=\dot{y}^2
\ee
is satisfied.
The expectation value is given in terms of supergravity as
\be \label{ww}
\langle W[C] \rangle =B e^{-\sqrt{\lambda}\tilde{I}},
\ee
where  the  prefactor $B$ has a
dependence on the loop $C$ which is subleading for large $\l$ and
$\tilde{I}$ is the Legendre transform of the worldsheet action
$I$ with respect to some of the loop variables \cite{dgo}.
The Legendre transform is needed because
some of the worldsheet scalars satisfy Neumann boundary conditions
instead of Dirichlet boundary conditions.
The area $I$ has a linear UV divergence $1/\e$
since the metric has a scale factor which
diverges as one goes near the boundary.
It was  demonstrated that
\cite{dgo}
the application of the Legendre transform removes
this UV divergence from the area and the result $\tilde{I}$  is
finite.

So far there has not been much discussions on the structure of the UV
divergences and their cancellation for Wilson loops in more general
gauge/gravity correspondence beyond the original  $AdS_5 \times
S^5$ case.  In a general supergravity
background where the metric is different from the simple $AdS_5 \times
S^5$ one, and where a nontrivial $B$-field and dilaton could be present,
there can be new kind of UV divergences. It is interesting to ask
whether the implementation of the Legendre transform can cure all the
UV divergences or not.
In \cite{myers}, the effects of a varying dilaton were
analysed by including the Fradkin-Tseytlin term for the dilaton
\cite{FT}.  It was
found that new UV-divergent terms proportional to $\sqrt{1/\e}$ and
$\log 1/\e$ occurs \footnote{
These divergences were computed for the worldsheet associated with the
Wilson line operator with fermion bilinear insertion. However it is
easy to see that these divergences are common to Wilson loop too.
}. Moreover these divergent terms cannot be subtracted away by
the application of Legendre transform.
A direct subtraction is applied to extract
a finite result. However, the subtraction  of the log-divergent term is
associated with a finite ambiguity and further physical input is
needed to fix the supergravity prediction for the
expectation value of the Wilson loop. This is unlike the cancellation
of the leading linear divergence
in the Polyakov action 
through a quadratic constraint on the loop variables, which has a nice
geometrical and physical interpretation.

In this paper, we focus on
the gravity dual analysis of the UV divergences 
from a nontrivial metric and $B$-field.
The main motivation of our work is to provide a general analysis of
the kind of UV divergence that may occur in the Wilson loop
correspondence and to  provide a prescription
for their cancellation.
We show indeed in general there are  new kinds of
UV divergences  associated with the metric and the  $B$-field
that cannot be cancelled away by the Legendre transformation. 
However, when certain  asymptotic conditions for
the metric and the $B$-field are satisfied,
the leading UV divergence becomes linear and one can cancel out the 
divergence with the Legendre transform by
choosing the open string boundary condition appropriately. Things are
different for the $B$-field. We find that the situation is similar to
the dilaton: in general the
divergences (if any) associated with the $B$-field cannot be
cancelled by the Legendre transformation.

Another motivation of this work is to  understand the role of
supersymmetry in the holographic correspondence of Wilson
loop in a general gauge/gravity duality. In the $\cN=4$ case, the
Wilson loop operator \eq{wilsono} preserves
some amount of local Poincare supersymmetry and is sometimes referred to
as ''locally BPS''. One may wonder if the finiteness of the Wilson
loop is related to the preservation of local supersymmetry.
Wilson loop operator, being a nonlocal divergent functional,
cannot be renormalized by the ordinary $R$-operation
\cite{IZ} restricted to the local operators. The renormalization
properties of Wilson loop with pure glue has been studied in, e.g.
\cite{renorm1,renorm2,renorm3}, and it was found that, apart from the
conventional wavefunction and coupling renormalization, the only
divergence in $W[C]$ is a factor $e^{-KL}$, where $K$ is a
regularization dependent linear
divergent constant and $L$ is the length of the loop. This is independent of the
form of $C$ and hence the Wilson loop is multiplicative
renormalizable. In $\cN=4$ SYM there is no wavefunction renormalization or
coupling renormalization, thus the finiteness of the expectation value
of the locally BPS Wilson loop means that the multiplicative
renormalization factor is finite. As is common in a supersymmetric
field theory, it is natural to associate the absence of renormalization
of this class of Wilson loop operators with the presence of local
supersymmetry, and to suspect that the later is responsible for it. It
is thus interesting to consider Wilson loop which preserves less or no
local supersymmetry and check if this is correct.

In the previous paper \cite{CG1}, we started to investigate this
question by considering the Wilson loop correspondence in the
Lunin-Maldacena duality \cite{LM}. The gauge theory is given by a
marginal $\b$-deformation of the $\cN=4$ SYM and has $\cN=1$
superconformal symmetries. Configuration of minimal surfaces that are
dual to field theory Wilson loop were constructed in \cite{naqvi}.
We proposed a form of Wilson loop operator that is the dual of these
string configurations. We also found
that, although these operators do not preserve any local supersymmetry,
they have finite expectation value (both in perturbation theory, which
we computed up to order $(g^2 N)^2$, and from supergravity).
In supergravity, the absence of divergence is due to  some special properties
satisfied by the metric and the  $B$-field. In field theory, we called
these operators ''near'' local BPS in order to distinguish them from generic
non-BPS Wilson loops whose expectation values are infinite
Although the operator is non-BPS, still there is the
possibility that  the cancellation of the UV divergence is due to
the underlying $\cN=1$ supersymmetric dynamics.

In this paper,
we find that the finiteness of the Wilson loop has nothing to do with
supersymmetry at all.
As in the $AdS_5 \times S^5$ case, the boundary constraint of the
worldsheet has an intermediate interpretation as a constraint on the loop
variables of the field theory Wilson loop operator.  It is a pure
coincidence that this loop constraint also implies a preservation of
local Poincare supersymmetry in the $\cN=4$ SYM theory. In general,
this condition has nothing to do with preservation 
of any supersymmetry.
In fact, as we will see, the multi-parameters $\b$-deformed
supergravity background \cite{frolov} is an example where the Wilson loop
expectation value is finite and where the background is not
supersymmetric.

The plan of the paper is as follows. In section 2, we present our
analysis of the UV divergence in the supergravity Wilson
loop associated with the $B$-field and the metric. In general the
divergence that may arises from the $B$-field coupling is of a different
structure from that in the Legendre transform and so cannot be
subtracted away. For background where such divergences are absent, the
leading order divergence arises from the area and it can be cancelled
away using Legendre transform if certain asymptotic conditions are
satisfied for the metric and the $B$-field and if the 
boundary coordinate of the open string satisfy a certain constraint.
As a consistency check, we show that this loop constraint guarantees 
that the loop equation is satisfied. Subleading divergences could be
present in general. We provide a stronger criteria on the supergravity background 
where the subleading divergences are absent and the
Wilson loop is expected to be finite. 
In section 3, we analyze the conditions for the cancellation of
leading divergence and show that they can be solved quite
generally. Some explicit backgrounds which satisfy these conditions are given
as examples.
Many of them also
satisfy the stronger form of the cancellation conditions and so for
these backgrounds, Wilson loop computed using the
supergravity description
\eq{ww} is finite.
As a final example, we consider the
Klebanov-Strassler background and show that
the leading linear divergence in the area can be cancelled away as
usual.
However there are
subleading divergences of order $(\log \e)^2$ associated with the
$B$-field and this cannot be cancelled away with the
Legendre transform. We end with some comments on the form of the Wilson loop
operator in the three dimensional  $\cN=6$ supersymmetric Chen-Simons theory  
of  Aharony,
Bergman, Jafferis and Maldacena \cite{ABJM} (ABJM).

\section{
Structures of UV divergence in the Wilson loop in general supergravity
background }

\subsection{Conditions on the supergravity background and the
string worldsheet for cancellation of leading order divergence}

Consider a general supergravity background. The string worldsheet is sensitive to
the metric, NSNS $B$-field and the dilaton. The structure of UV
divergence associated with a varying
dilaton has been analysed in \cite{myers} and we will focus on
analysing the effect of a general metric and transverse $B$ field on the UV
divergences of the supergravity Wilson loop.
Denote the metric in the
string frame as
\be \label{metric1}
ds^2  =G_{\m\n} dX^\m X^\n + G_{ij} dY^i dY^j,
\ee
where $\m,\n =1,\cdots, m$ denotes the indices of a $m$-dimensional
spacetime; and $i,j
=1, \cdots, n$ denotes the indices of a $n$-dimensional internal
manifold. For this metric to be relevant for a holographic correspondence,
we assume
that the metric has a (conformal) boundary at $Y =0$,
where $Y := \sqrt{(Y^i)^2}$ is the
radial variable and is of length dimension.
It is also convenient to introduce the angular variables
$\th^i$ where
$Y^i =Y \theta^i$ with $\th^i{}^2 =1$.
We will  assume that in the leading order in $Y$, the metric
have the following asymptotic dependence near the
boundary:
\be \label{cond-asy}
G_{\m\n} = \frac{h_{\m\n}}{Y^{\a}} + \cdots,
\qquad G_{ij} = \frac{k_{ij}}{Y^{\b}} + \cdots, \quad \mbox{as $Y \to 0$}
\ee
for $\a,\b \geq 0$.
Here
$h_{\m\n}, k_{ij}$ are functions of $\th^i$ only and    $\cdots$ denotes
subleading terms.

Next let us analyze the string boundary condition.
Let $(\s_1,\s_2) =(\t,\s)$ be the worldsheet coordinates.
The worldsheet action  of the string is
\be
I = \int_\S d^2 \s  ( \sqrt{\det g} - i B_{ij} \del_1 Y^i \del_2 Y^j ) ,
\ee
where $g_{\a\b} = G_{IJ} \del_\a X^I\del_\b X^J$ is the  induced
metric. We note that since the worldsheet is an open one, the $B$
field coupling itself is not invariant under the gauge transformation
$\d B = d \L$. In order to be gauge invariant,  the $B$  term
should be supplemented with a boundary coupling
$\int_{\del \S} \cA$. Without writing
this term, we are assuming we are in a gauge where $\cA=0$ and $B$ is
the corresponding potential in this gauge. However how to fix this
choice of $B$-field is a subtle issue. Similar subtlety also arise in
the computation of Wilson loop expectation value using D3-brane
dual where one need to know the form of the RR 4-form potential $C_4$ used
in the WZ coupling of the D3-brane \cite{df}. There a symmetry criteria
is used to pick a certain natural form of $C_4$. We will assume that
similar considerations can be applied and the
correct form of $B$ field is used in the analysis below.

The equation of motion implies the Hamilton-Jacobi equation
\bea\label{HJ0}
G^{ij}(P_i -  i B_{ik} \partial_1 Y^k) (P_j -  i B_{jl} \partial_1
Y^l) + G^{\m\n} P_\mu P_\nu
 = G_{ij} \partial_1 Y^i \partial_1  Y^j+
 G_{\mu\nu}\partial_1 X^\mu \partial_1  X^\nu
\eea
where
\be
P_{i}= G_{ij} J_1{}^\b \partial_{\b}Y^{j} + i B_{ij} \partial_1 Y^j, \quad
P_\mu =  G_{\m\n}J_1{}^\b \partial_\b X^\nu. \label{moment}
\ee
are the momentum and
\be
J_\a{}^\b = \frac{1}{\sqrt{g}} g_{\a\g}\e^{\g\b}
\ee
is the complex structure ($\a,\b =1,2$) on the worldsheet.
Substitute the conjugate momentum,
we obtain
\bea\label{hj}
\frac{k_{ij}}{Y^{\b-\a}} J_1{}^\a \partial_{\a}Y^{i} J_1{}^\b \partial_{\b}Y^{j}
+  h_{\m\n} J_1{}^\a \partial_{\a}X^{\m} J_1{}^\b    \partial_{\b}X^{\n}
 =\frac{k_{ij}}{Y^{\b-\a}} \partial_1 Y^i \partial_1  Y^j+
 h_{\m\n} \partial_1  X^\m \partial_1  X^\n \nn\\
{}
\eea
near $Y=0$.

One like to know how this equation put constraint on the
boundary variables of the theory.
To do this we need the boundary conditions for  the string coordinates.
Suppose that the Wilson loop is parametrized by
$(x^\mu(\sigma_1),y^i(\sigma_1))$ and choose the world-sheet coordinates
such that the boundary is located at $\sigma_2=0$. First we have the
Dirichlet boundary condition for the coordinates
\be
X^\mu(\sigma_1,0)=x^\mu(\sigma_1). \label{dirichlet}
\ee
For the remaining coordinates $Y^i(\sigma_1,\sigma_2)$,
due to the presence of the $B$-field, we 
propose the mixed
boundary condition
\bea\label{bci}
J_1^\a \partial_\a Y^k(\sigma_1,0)
+ i B^{k}{}_{l} \partial_1 Y^l(\sigma_1,0) =
 E^k{}_{l}\, \dot{y}^l(\sigma_1),
\eea
where $E^k{}_{l}$ is some invertible matrix which can depend on 
$Y,\,\th^i$. Its form will be determined later.

For now, focus on  the  first term on  the RHS of \eq{hj}.
For a string which
terminates at the boundary, it is $Y^i(\s_1,0) =0$. This would imply
also $\del_1 Y^i(\s_1,0) =0$. If $\b -\a \leq 0$, then we can get rid
of this term immediately. If $\b -\a > 0$, then this
term indeterminate. To proceed, we consider a
limiting process of letting $Y\to 0$. One can get rid of this term if
\footnote{
We use the symbol $f=o(g)$  to mean $\lim f/g =0$, i.e. $f$ tends to
infinity slower than $g$ or $f$ tends to
zero faster than $g$.
We also use  $f=O(g)$ to mean $\lim f/g =k$, $ 0 \le k < \infty$.
i.e.  $f$ tends to infinity not faster than $g$ or $f$ tends to
zero not slower than $g$ or $f$ tends to infinity not faster than $g$.
}
$ \partial_1 Y^i = o(Y^{\frac{\b-\a}{2}})$.
As in the $AdS_5 \times S^5$ case,
the term $h_{\m\n} J_1{}^\a \partial_\a X^\mu J_1{}^\a \partial_\a
X^\nu$
on the 
LHS of \eq{hj}
has to vanish near a smooth boundary since otherwise 
the determinant of the induced metric will blow up and this will
cost an infinite
area. Therefore we arrive at the condition
\be \label{bc2}
h_{\m\n} \dot{x}^\m \dot{x}^\n =
 \frac{1}{Y^{\b-\a}}
k_{ij} J_1{}^\a \partial_{\a}Y^{i} J_1{}^\b \partial_{\b}Y^{j}
\ee
for a worldsheet which terminates on the boundary $Y=0$. In order for
the condition to make sense, one need $J_1{}^\a \partial_{\a}Y^{i}$ to
be of the order of $Y^{\frac{\b-\a}{2}}$.

Before  analysing further the boundary condition, let us
turn to an analysis of the divergence
in the worldsheet action $I$ and its
Legendre transform
\be
\tilde I =I-\oint d\sigma_1 P_i Y^i.
\label{naction}
\ee
As in the $AdS_5 \times S^5$ case, the area $A$ may  pick up
a divergent contribution from the boundary. This can be seen by writing
the metric in the form
\be
G_{ij} dY^i dY^j= \frac{k_{ij} \th^i \th^j}{Y^{\b}} dY^2
+ \frac{1}{Y^{\b-2}} k_{ij} d\theta^i d\theta^j + \frac{2}{Y^{\b-1}}
k_{ij} \th^i d \th^j dY + \cdots,
\ee
where $\cdots$ denotes terms coming from the subleading expansion terms in the
metric \eq{cond-asy}.
Near the boundary, $A$ picks up the dominant contribution 
\be
\int dY\, d\s_1 \frac{\sqrt{k_{ij}\th^i\th^j}}{Y^{\frac{\a+\b}{2}}}\sqrt{h_{\m\n}
  \dot{x}^\m\dot{x}^\n} + \cdots.
\ee
Since the metric is singular at $Y=0$, we
introduce a regulator $Y=\e$ and  evaluate the
regularized action for $Y\geq\epsilon$.
The divergent part of the area is
\be \label{div1}
A = \frac{c}{\e^{(\a+\b)/2-1} }\int d\s_1\sqrt{ k_{ij}\th^i\th^j}
\sqrt{h_{\m\n}
  \dot{x}^\m\dot{x}^\n} + \cdots,
\ee
where $c^{-1}:=(\a+\b)/2-1$ and $\cdots$ denotes possible
subleading divergent terms. The $B$-field coupling can be written as
\be
-i \int B_{ij} \del_1 Y^i \del_2 Y^j  = -i \int \del_2 (B_{ij} \del_1 Y^i Y^j) +
i \int \del_2(B_{ij} \del_1 Y^i) Y^j  .
\ee
With the cutoff $Y=\e$, the first term on the RHS contributes
the boundary term
\be
\oint d\s_1  i B_{ij} Y^i \del_1 Y^j \big|_{Y=\e},
\ee
which cancels against the $B$-dependent term from the
Legendre transform
\be \label{PY}
P_i Y^i = G_{ij} Y^i J_1{}^\a \del_\a Y^j + i B_{ij} Y^i \del_1 Y^j.
\ee
Therefore we can write
\be
\tilde I = \tilde I_A + \tilde I_B,
\ee
where
\be\label{tilde-A}
\tilde I_A := A - \oint d \s_1 \; G_{ij} Y^i J_1{}^\a \del_\a Y^j ,
\ee
\be \label{tilde-B}
\tilde I_B := i \int d^2 \s \; \del_2 (B_{ij} \del_1 Y^i) Y^j,
\ee
are the Legendre transform modified contributions of the area and
$B$-coupling term.
There is a reason we group the terms in this way.
Note that the term $G_{ij} Y^i J_1{}^\a \del_\a Y^j$ is of the order
of $1/Y^{\frac{\a+\b}{2} -1}$ and is of precisely the same order of
divergence as in $A$. Note also that
$A$ has a dependence in $J_1{}^\a \del_\a Y^j$ due
to \eq{bc2}. Thus  it is 
in principle 
possible to cancel the divergence in $A$ using
the term $\oint G_{ij} Y^i J_1{}^\a \del_\a Y^j$.
On the other hand, the term $\tilde I_B$ depends on $\del_1 Y^i$. This
dependence is different from the other terms.
Thus the $B$-field contribution, if divergent,
corresponds to a new divergence with a different type of functional
dependence on the variables of the theory.

Let us consider a $B$-field such that
\be \label{cond-B}
 B_{ij} \del_1 Y^i =
o\left(\frac{1}{Y^{\frac{\a+\b}{2}}}\right).
\ee
This implies that the divergence in $\tilde I_B$ will
be subleading compared to $\tilde I_A$.
This condition also implies that the second term on the 
LHS of
\eq{bci} behaves asymptotically as
\be \label{BY}
i B^k{}_l \del_1 Y^l = o(Y^{\frac{\b-\a}{2}}).
\ee
Since  $J_1{}^\a \del_\a Y^k$ is the order of
$Y^{\frac{\b-\a}{2}}$, one can drop the $B$-term in \eq{bci}.
It is convenient to define $E^k{}_l = Y^{\frac{\b-\a}{2}}
\L^k{}_l$ and the
boundary condition \eq{bci} can be written as
\be \label{bc-final}
J_1^\a \partial_\a Y^k(\sigma_1,0)  =
Y^{\frac{\b-\a}{2}} \L^k{}_{l}\, \dot{y}^l(\sigma_1) .
\ee
The Hamilton-Jacobi equation \eq{bc2} becomes
\be\label{gen-loop}
h_{\m\n} \dot{x}^\m \dot{x}^\n =
k_{ij} \L^i{}_m \L^j{}_n \dot{y}^m \dot{y}^n.
\ee
This condition will play a key role in the cancellation of the divergences
in $\tilde I_A$. To see this, note that
\be \label{l1}
G_{ij} Y^i J_1{}^\a \del_\a Y^j =
  \frac{1}{Y^{\b-1}} k_{ij} \th^i \th^j  J_1{}^\a \partial_\a Y +
\frac{1}{Y^{\b-2}} k_{ij} J_1{}^\a \th^i \partial_\a \th^j + \cdots ,
\ee
where $\cdots$ denotes the subleading contribution from the asymptotic
expansion
of the metric \eq{cond-asy}.
This is to be compared with the leading divergence
$\sqrt{k_{ij} \th^i \th^j }$ 
$\cdot$ $\sqrt{h_{\m\n}\dot{x}^\m\dot{x}^\n}/ Y^{\frac{\a+\b}{2}-1}$ in $A$,
which, using \eq{bc2}, can be
written as follows:
\be\label{l2}
\frac{\sqrt{k_{ij} \th^i \th^j }}{Y^{\b-1}}
\sqrt{
(J_1{}^\a\del_\a Y)^2 k_{ij} \th^i \th^j
+ 2 Y  J_1{}^\a \del_\a Y J_1{}^\b k_{ij} \th^i \del_\b \th^j
+ Y^2 J_1{}^\a J_1{}^\b k_{ij} \del_\a \th^i \del_\b \th^j
}.
\ee
Obviously \eq{l1} and \eq{l2} cannot match in general. Doing so will
require an extra constraint among the derivatives of $\th^i$ and
$Y$, which, first of all, is not obvious it is in consistent with the relation
\eq{bc2}. Moreover this relation does not have any obvious physical
interpretation in field theory. On the other hand there is a
particularly simple set of conditions which guarantee that
\eq{l1} and \eq{l2} are equal, namely,
\bea
k_{ij} \th^i = \th^j, \label{cond-g} \\
\b -\a <2 \label{cond-g2}.
\eea
In fact the first condition implies immediately $k_{ij} \th^i \del_\a
\th^j =0$ and  hence the vanishing of the second term
in \eq{l1} and \eq{l2}; while the second
condition says that the last term in \eq{l2} is subleading compared to
the first term.
As a result of \eq{cond-B}, \eq{cond-g} and \eq{cond-g2}, we can write
\be
G_{ij} Y^i J_1{}^\a \del_\a Y^j =
  \frac{1}{Y^{\b-1}} J_1{}^\a \partial_\a Y +\cdots
= \frac{1}{Y^{\b-1}}
\sqrt{ k_{ij} J_1{}^\a \partial_\a Y^i J_1{}^\a \partial_\a Y^j } + \cdots
\ee
near $Y =0$, and the Legendre transform contributes the
singular terms
\be \label{div2}
\oint d\sigma_1 P_i Y^i
= \frac{1}{\e^{(\a+\b)/2-1}} \oint d\sigma_1
\sqrt{k_{ij} \Lambda^{i}{}_{m}\Lambda^{j}{}_{n}\,
\dot{y}^m \dot{y}^n}
+ \cdots,
\ee
where we have used \eq{bc-final}.
Therefore the leading divergence term in \eq{div1}, \eq{div2} cancels
if $c=1$, i.e. if the leading divergence is linear:
\be \label{A-tilde}
\tilde{I}_A =\frac{1}{\e}\oint \left (
\sqrt{h_{\m\n} \dot{x}^\m\dot{x}^\n}
-\sqrt{ k_{ij} \Lambda^{i}{}_{m}\Lambda^{j}{}_{n}\, \dot{y}^m
  \dot{y}^n}
\right )
+ \cdots,
\ee
and  if the Hamilton-Jacobi condition \eq{gen-loop} holds.
Here $\cdots$ denotes the subleading contribution from the asymptotic
expansion of the metric \eq{cond-asy}. Whether
there are further subleading singularity (like, for example,
$1/\sqrt{\e}$
or $\log \e$ type)  or not will depend on the specific details of the
asymptotic form of the
background metric.
Note that since $\del_1 Y^i $ is of order $Y$,
the sufficient condition \eq{cond-B} for the $\tilde I_B$-term to be
subleading divergent   can be written as
\be \label{cond-B1}
 B_{ij} =
o \left( \frac{1}{Y^{\frac{\a+\b}{2}+1}} \right).
\ee
On the other hand, if
\be \label{cond-B2}
 B_{ij} =
o\left( \frac{1}{Y^2} \right),
\ee
then the $\tilde I_B$-term is non-divergent.

Summarizing in a general supergravity background, the $B$-field
coupling in the worldsheet action generically generates a divergence
which cannot be cancelled with the Legendre transform. A sufficient
condition for the $B$-field contribution to be finite is \eq{cond-B2}.
When there is no such divergence, the leading order
divergence in the Wilson loop arises from the area and it
can be cancelled with the application of
Legendre transform if  the following
conditions are satisfied:

\begin{enumerate}

\item supergravity background:

- The supergravity metric takes the asymptotic form
\eq{cond-asy} near the boundary. Moreover
\be \label{ab}
\a +\b =4, \quad \b -\a <2.
\ee

- The boundary metric $h_{\m\n}$ is independent of $\th^i$. The
transverse part of the metric satisfies the boundary
condition
\be
k_{ij} \th^i = \th^j.
\ee

These conditions  are conditions on the background
and do not impose any extra constraint on the form of the Wilson loop
variables.

\item string worldsheet:

The boundary constraint \eq{gen-loop} for the string worldsheet
is  satisfied.

\end{enumerate}

In general, once the leading UV divergences are cancelled,
there may be further subleading singularity (like, for example,
$1/\sqrt{\e}$  or $\log \e$ type).  An extensive analysis of them will
need information on the specific details of the
asymptotic form of the
background metric, the $B$-field and the dilaton. Generally we don't
expect the subleading
divergences can be cancelled with the application of Legendre transform.

A special situation with no further
subleading divergence is if the leading correction term in the
asymptotic conditions  \eq{cond-asy} and \eq{cond-B2}
are of at least order $Y$. We will examine some examples of this kind
later.

\subsection{Comments: boundary constraint as  loop constraint}

Just as in the original $AdS_5 \times S^5$ case, one would like to
interpret the boundary constraint \eq{gen-loop} for the open string
as a condition  in the field theory.
Since the Wilson loop is specified by the loop variables
$\dot{x}^\mu$ and $\dot{y}^i$, and $\th^i$ does not play any role, 
the loop constraint should not depend on $\th^i$.
This means $h_{\m\n}$ should be independent of  $\th^i$. 
For the same reason, one should choose
${\L}^k{}_m$ such that $k_{kl} {\L}^k{}_m {\L}^l{}_n$ is
independent of $\th^i$. Generally this can be achieved  by taking
$\L^k{}_m$ of the form
\be
\L^k{}_m = \hat{\L}^k{}_l M^l{}_m,
\ee
where $\hat{\L}^k{}_l$  is the vielbein  of the metric $k_{kl}$ and $M^l{}_m$
is an invertible matrix which is independent of $\th^i$ but can
depends arbitrarily on parameters which have
meaning both in supergravity and in the field theory (e.g. the
't Hooft coupling or parameters in the theory such as the $\b$-deformation
parameter in the Maldacena-Lunin duality).
As a result, the condition \eq{gen-loop} takes the form
\be \label{constraint-final}
h_{\m\n} \dot{x}^\m \dot{x}^\n =
a_{ij}\dot{y}^i \dot{y}^j, \qquad i,j = 1,\cdots, n
\ee
where we have defined $a_{ij} := M^n{}_i M^n{}_j$. In general
the form of the matrix $a_{ij}$ will be a function of the couplings of
the theory and cannot be fixed from the supergravity
analysis alone.
In the original $\cN=4$ SYM case \cite{dgo}, the matrix $a_{ij}$ is
given by $a_{ij} =\d_{ij}$.
We have also computed the constraint
for the $\cN=1$ $\b$-deformed superconformal field theory
and find $a_{ij}=\d_{ij}$ up to $\l^2$ order  in perturbation theory \cite{CG1}.
We emphasize that in general
the constraint \eq{constraint-final}
has nothing to do with preservation of any
supersymmetry. It is a pure
coincidence that this loop constraint also implies a preservation of
local Poincare supersymmetry in the $\cN=4$ SYM theory.

Let us make a consistency check on the boundary constraint
\eq{constraint-final}.
In the large $N$ limit of  gauge theory,  Wilson loop satisfies a
closed set of equations called the loop equation
\cite{loopeqn}.  To further justify the supergravity procedure for the
computation of the Wilson loop expectation value,
one should check that the supergravity ansatz\eq{ww}
satisfies the loop equation \cite{loopeqn}.
As in the $AdS_5
\times S^5$ case, although the leading linear divergence cancels out
when the loop constraint \eq{constraint-final} is satisfied, the loop
variation does not commute with the constraint and so the linear divergence
may gives a divergent contribution and violate the loop equation. We
show this is not the case.

The loop derivative operator is given by
\be\label{loop-deriv}
\hat L=\lim_{\eta\rightarrow 0}
\oint ds \int_{s-\eta}^{s+\eta} ds'
\left( \frac{\delta^2}{\delta x^\mu(s')\delta x_\mu(s)}
-a^{ij} \frac{\delta^2}{\delta y^i(s')\delta y^j(s)}\right).
\ee
That this definition is correct  can be confirmed by checking that
$\hat L \langle W \rangle =0 $ in field theory for the  Wilson loop
operator \eq{wilsono}.
As usual the loop regulator $\eta$
has to be taken much smaller than the
UV cutoff scale $\e$ in order to extract the equation of motion terms.
Now acting on the supergravity ansatz 
\eq{ww} with the the loop operator, we get the
leading term in  large $\l$,
\be
\lambda \lim_{\eta\rightarrow 0}
\oint ds \int_{s-\eta}^{s+\eta} ds'
\left(\frac{\delta\tilde I_A}{\delta x^\mu(s')}
\frac{\delta\tilde I_A}{\delta x_\mu(s)}
-\frac{\delta\tilde I_A}{\delta y^i(s')}
\frac{\delta\tilde I_A}{\delta y_i(s)}\right).
\label{leadingloop}
\ee
Let us now extract the divergent contribution from $\tilde{I}_A$ 
in \eq{A-tilde}.
Given the condition \eq{constraint-final}, we can
choose a parametrization such that $h_{\m\n} \dot{x}^\m\dot{x}^\n =
a_{ij} \dot{y}^i \dot{y}^j=1$ and get
\be
\hat L \langle W \rangle =
\frac{\lambda \eta}{\epsilon^2}
\oint ds
\left(h_{\m\n} \ddot x_\mu \ddot x^\n
-a_{ij}\ddot{y}^i\ddot{y}^j)\right).
\label{leadingloop1}
\ee
For a smooth loop the terms in the integral are finite.
Therefore by taking  $\eta$ going  to zero faster than
$\epsilon^2$,  we find
\be
\hat L \langle W \rangle =0
\ee
and the loop equation is satisfied.

\section{General solution to the conditions on SUGRA background and
  examples }

\subsection{General solution to the metric condition}

The condition \eq{cond-g} on the metric may look
a little  restrictive at first sight. We show now that
it is in fact satisfied by a
general class of metric of the form
\be \label{gen-metric}
ds^2 = {H}_1(Y) dT^2 +{H}_2(Y) d \vec{X}^2 +{F}(Y) d
Y^2 + g_{ij} d \theta^i d \theta^j,
\ee
where $\theta^i,\,i,j=1, \cdots, n$ are the
coordinates of the $n-1$ dimensional space $X_{n-1}$; and the metric
$g_{ij}$  is a function of $Y^i$, e.g. as in the Klebanov-Strassler
metric \cite{KS}.
The metric can be thought as a warped product of
the boundary spacetime $(T,\vec{X})$
and the transverse space $(Y,\th^i)$.

Defining $Y^i= Y \theta^i$ and making the coordinate transformation we get
\be \label{3}
g_{ij} d\theta^i d\theta^j=\frac{1}{Y^2} (g_{kl}+ g_{ij}\theta^i
\theta^j \theta^k \theta^l  - g_{il} \theta^i \theta^k- g_{ki}
\theta^i \theta^l) dY^l dY^k.
\ee
So our metric become
\bea \label{metricfinal}
ds^2 = H_1(Y) dT^2 +H_2(Y) d \vec{X}^2 +G_{ij} dY^i dY^j ,
\eea
where
\be
G_{ij} := F(Y) \theta^i\theta^j+\frac{1}{Y^2}
A_{ij},\label{gkl}
\ee
and
\be \label{Akl}
A_{ij}:= g_{ij}+ g_{kl}\theta^k \theta^l \theta^i \theta^j  -
g_{il} \theta^l \theta^j- g_{jl} \theta^l \theta^i.
\ee
The matrix $A_{ij}$ satisfies the following identity,
\be\label{Aident}
A_{ij}\theta^j=0,
\ee
and so
\be \label{GF}
G_{ij} Y^j = F(Y) Y^i
\ee
Note that \eq{GF} is of the form of \eq{cond-g}. Therefore if
$F$ behaves as
\be
F(Y) = \frac{1}{Y^\b}, \quad Y \to 0,
\ee
near the boundary, then the condition \eq{cond-g} is
satisfied. Therefore if also $\a+\b=4$ and $\b-\a <2$,
then the metric conditions are satisfied.

It is easy to give example where the condition \eq{cond-g} is not
satisfied. For example, if we have started with a metric with an
additional cross-terms $dY d \th^i$
\be \label{metric2}
ds^2 = H_1(Y) dT^2 +H_2(Y) d \vec{X}^2 +F(Y) d Y^2 +K_i(Y)dY
d\theta^i+ g_{ij} d \theta^i d \theta^j,
\ee
then under the
same coordinate transformation, the additional term takes the form
\be
K_i(Y)dY d\theta^i =
\frac{1}{Y} \Big(\frac{1}{2}(\theta^k K_l+\theta^l K_k)- (K_i \th^i) \theta^k
\theta^l\Big)dY^k dY^l
:= \frac{1}{Y} \xi_{kl} dY^k dY^l .
\ee
$\xi_{kl}$ satisfies the following identities
\be
\xi_{ij} \th^j = \frac{1}{2}( K_i -(K_l\th^l) \th^i), \qquad
\xi_{ij}\theta^i \theta^j=0,\qquad
\xi_{ij}\theta^i \partial\theta^j= \frac{1}{2} K_l\partial\theta^l
\ee
Denote the whole metric as $G_{ij} :=H_{ij}+Y^{-1} \xi_{ij}$, where $H_{ij}$
is given by the RHS of \eq{gkl}. It is
\be
G_{ij} \th^j = F(Y) \th^i + \frac{1}{2Y}( K_i -(K_l\th^l) \th^i).
\ee
Since the right hand side is generally not proportional to $\th^i$,
the condition \eq{cond-g} is no longer satisfied.
Note that the cross-terms in \eq{metric2} may be
eliminated with a shift of $\th^i \to
\th^i + a_i(Y)$. However the new $\th$'s will
not satisfy the condition $(\th^i)^2 =1$ anymore. This is another way
to see that the metric conditions are not satisfied.

\subsection{Examples}

Here we examine some backgrounds with known dual field theories,
to which our analysis can be applied.\\

$\bullet$ \textbf{Background with $AdS_5 \times X^5$ metric }\\
This is a standard example. The metric of the space can be written  as
\be
ds^2=U^2 \sum_{\mu=0}^3 dX^\mu dX^\mu + \frac{dU^2}{U^2}+ d X_5^2.
\ee
where  $X^5$ is an internal compact space.
In this case $\a=2=\b$ and the condition \eq{ab} is satisfied.
The linear divergence in $A$ is cancelled by the Legendre transform
and $\tilde I_A$ is finite.
Some explicit examples are,
$X^5 =
 S^5$, $\tilde{S}^5$, $\tilde{S}_{\g_1,\g_2,\g_3}^5$,
$T^{1,1}$, $Y^{p,q}$, $L^{p,q,r}$,
etc.,
where respectively these spaces are the 5-sphere for the original Maldacena
AdS/CFT correspondence \cite{Mal}, the $\b$-deformed 5-sphere for the
Lunin-Maldacena $\b$-deformation \cite{LM},
the multi-parameter $\b$-deformed sphere \cite{frolov},
and the Sasaki-Einstein spaces \cite{SE1,SE2}.
The boundary
condition for the string
minimal surface is
\be
J_1^\a \partial_\a Y^k(\sigma_1,0)=
\hat{\Lambda}^k{}_{m} M^m{}_l \, \dot{y}^l(\sigma_1).
\ee

It is easy to see that $\tilde I_B$ is finite for
these cases. In the  $AdS_5 \times S^5$
case or in the duality with Sasaki-Einstein spaces, there is simply
no $B$-field.
In the $\b$-deformation  or the
multi-parameters $\b$-deformation, the $B$-field is of the form
\be \label{gen-B}
B = \frac{1}{2}B_{ab} d \phi^a d \phi^b,
\ee
where  $\sum (\mu_a)^2 =1$,
$\phi^a$ $(a=1,2,3)$ are the azimuth angles defined by
\bea
&&Y^1 =Y \theta^1 =Y \mu_1 \cos\phi_1,\,\, \quad
Y^4= Y \theta^4 = Y \mu_1 \sin\phi_1,\nn\\
&&Y^2 = Y\theta^2= Y \mu_2 \cos\phi_2,\,\, \quad
Y^5 = Y\theta^5= Y \mu_2 \sin\phi_2,
\label{yy}\\
&&Y^3 = Y\theta^3 = Y \mu_3 \cos\phi_3,\,\,\quad
Y^6 =Y\theta^6 = Y  \mu_3 \sin\phi_3\nn
\eea
and $B_{ab}$ is a function of $\mu_a$. This form of the $B$-field
respects the symmetries of 
the $\b$-deformed sphere and we will take it to be the
$B$-field where the string is coupled to. In general one may get a
different answer by using a different gauge equivalent $B$-field. This
is similar to the situation discussed in \cite{df} where an open
D3-brane is employed to compute the expectation value of Wilson loop in
higher representation. There the answer is shown to depend on the
gauge choice of the  RR 4-form potential $C_4$ which appears in the 
Wess-Zumino couping. A symmetry argument was used to suggest the
natural form of the $C_4$ to be used. 

Obviously the $B$-term in the worldsheet action is finite. For the
piece $B_{ij} Y^i \del_1 Y^j$ in the Legendre transform, since
$B_{ij}$ is of order $1/Y^2$, this term is potentially linear
divergent. However this does not happen since, as we have shown in
\cite{CG1}, a $B$-field of the form \eq{gen-B} satisfies the
condition
\be
B_{ij} Y^i =0
\ee
exactly. This can be seen easily by noticing that
\bea
d\phi^1 d \phi^2 =
\frac{1}{\m_1^2 \m_2^2 Y^4 }
(Y_4 Y_5 d Y_1\wedge d Y_2 + Y_1 Y_2 d
Y_4\wedge d Y_5 + Y_1 Y_5 d Y_2\wedge d Y_4 - Y_2 Y_4 d
Y_1\wedge d Y_5), \nn\\
d\phi^1 d \phi^3
= \frac{1}{\m_1^2 \m_3^2 Y^4 }
(Y_4 Y_6 d Y_1\wedge d Y_3 + Y_1 Y_3 d Y_4\wedge d Y_6
+ Y_1 Y_6 d Y_3\wedge d Y_4 - Y_3 Y_4 d Y_1\wedge d Y_6), \nn\\
d\phi^2 d \phi^3=
\frac{1}{\m_2^2 \m_3^2 Y^4 }
(Y_5 Y_6 d Y_2\wedge d Y_3 + Y_2 Y_3 d Y_5\wedge d Y_6
+ Y_2 Y_6 d Y_3\wedge d Y_5 - Y_3 Y_5 d Y_2\wedge d Y_6). \nn
\eea
As a result, the piece $B_{ij} Y^i \del_1 Y^j$ in the Legendre transform
is zero. Therefore, there is no divergence associated with the
$B$-field. This can also be checked using \eq{tilde-B}. For example
the contributions from $B_{12}, B_{15}$ to $\del_2(B_{ij} \del_1 Y^i)
Y^j$ is of the form
$ \sim
\frac{Y_4 (Y_2)^2}{Y^4} \del_1 Y_1  \del_2 \frac{Y_5}{Y_2} .
$
This is finite as $Y\to 0$ and so $\tilde I_B$ is free from any divergence.
Also
since there is no subleading correction terms to the metric and the $B$-field,
there is
no subleading divergence at all. The Wilson loop is finite.

We remark  that the background $AdS_5 \times
\tilde{S}_{\g_1,\g_2,\g_3}^5$ for the multi-parameters $\b$-deformation
is not supersymmetric, but the Wilson loop expectation value is
finite. This clearly shows that
supersymmetry or  the satisfaction of the BPS condition for the loop
is not what is required for the finiteness of Wilson loop expectation
value.

$\bullet$
\textbf{Supergravity background with asymptotically
$AdS_5 \times X^5$ metric}\\
The first kind of example is given by a finite temperature deformation
of any of the metric above. For example for $\cN=4$ at finite
temperature, the metric  is
\be\label{fintemp}
ds^2=U^{2}\Big(-(1-\frac{U_T^4}{U^4})dt^2+
(dX^i)^2\Big)+\Big(1-\frac{U_T^4}{U^4}\Big)^{-1}\frac{dU^2}{U^2}+
d\Omega_5^2
\ee
Asymptotically, the metric behaves identically to  that of the $AdS_5 \times S^5$
background.
So the cancellation of the infinity occurs
with the same boundary conditions as in the $AdS_5 \times S^5$ case.
Putting a finite temperature deforms the asymptotic
form of the metric with power-like terms and this does not introduce
any additional subleading singularity.

$\bullet$ \textbf{Sakai-Sugimoto QCD model}\\
The background consists of a dilaton, a RR 3-form potential and
the metric  \cite{SS}
\bea\label{d4}
ds^2 &=& \left (\frac{U}{R} \right )^{3/2} (\eta_{\m\n} dX^\m dX^\n+ f(U)dz^2)
+\left (\frac{R}{U} \right )^{3/2}\left(
\frac{dU^2}{f(U)}+U^2 d\Omega_4^2 \right), \nn\\
e^\phi &=& g_s \left( \frac{U}{R}\right)^{3/4}, \qquad
f(U) = 1-\frac{U_{KK}^3}{U^3}.
\eea
Here $X^\m$ ($\m=0,1,2,3$) is the spacetime.
$z=X^5$ is periodic and describes the compact direction of the D4-brane.
$U >U_{KK}$ corresponds to the radial direction transverse to the
D4-brane.
With the coordinate transformation $Y=R^2/U$, the metric
near the boundary $U=\infty$ reads
\be
ds^2=\left(\frac{R}{Y}\right)^{3/2}
(\eta_{\m\n} dX^\m dX^\n+ d z^2)
+ \left(\frac{R}{Y}\right)^{5/2}
(dY^2+Y^2 d\Omega_4^2).
\ee
In this case $\a=3/2, \b=5/2$ and the condition \eq{ab} is
satisfied. The  leading UV divergence is a linear one and it can be
cancelled with a choice of the boundary
condition for the string
minimal surface
\be
J_1^\a \partial_\a Y^k(\sigma_1,0)= Y^{1/2} M^k{}_l \, \dot{y}^l(\sigma_1).
\ee
The  vielbein is trivial since $k_{ij} =\d_{ij}$ ($i,j=1,\cdots, 5$) 
for the boundary
metric. Including the contribution of the pion field $\varphi_0$, we
propose the following form of the Wilson loop
operator for the Sakai-Sugimoto QCD model,
\be
W[C] =\frac{1}{N}\Tr \,P\,
\exp\left(\oint_C d\t (iA_\mu \dot x^\mu+ i \varphi_0 \dot z
+ \varphi_i\dot y^i)\right),
\ee
and the constraint is
\be
\dot{x_\m}^2 = \dot{y_i}^2 -\dot{z}^2.
\ee
Moreover since the subleading correction terms to the
metric is power-like, therefore there is no further subleading
UV divergences.

$\bullet$ \textbf{Klebanov-Strassler background}\\
Another example is the Klebanov-Strassler background \cite{KS}
which describes a warped deformed conifold.
In this case
the asymptotic behavior of the metric is different from the power
ansatz \eq{cond-asy}. However it is not difficult to
repeat our analysis above.

The background has a  constant dilaton, a RR 2-form, and the metric
and $B$-field
\bea
ds^2&=&h^{-1/2}m^2 dx_m dx_m+h^{1/2}\frac{3^{1/3}}{2^{4/3}}K\left[
\frac{1}{3K^3}(d\tau^2+(g_5)^2)+\cosh^2\frac{\tau}{2}[(g_3)^2+(g_4)^2]
\right.\nn\\
&&\left. \qquad \qquad \qquad \qquad \qquad \qquad\qquad \qquad \qquad +
\sinh^2\frac{\tau}{2}[(g_1)^2+(g_2)^2]
\right], \label{KS-metric} \\
B&=&\frac{g_s M}{2}\left[f g_1\wedge g_2+k g_3\wedge g_4\right],
\label{KS-B}
\eea
where
$g_i$ is a basis of invariant one-form on $T^{1,1}$
\bea
&&
g_1=\frac{1}{\sqrt{2}}(-s_1 d\phi_1-c_\psi s_2 d\phi_2+s_\psi d\theta_2),
\qquad
g_2=\frac{1}{\sqrt{2}}(d\theta_1-s_\psi s_2 d\phi_2-c_\psi d\theta_2),
\nonumber\\
&&
g_3=\frac{1}{\sqrt{2}}(-s_1 d\phi_1+c_\psi s_2 d\phi_2-s_\psi d\theta_2),
\qquad
g_4=\frac{1}{\sqrt{2}}(d\theta_1+s_\psi s_2 d\phi_2+c_\psi d\theta_2),
\nonumber\\
&&g_5=d\psi+c_1 d\phi_1+c_2 d\phi_2.
\eea
The $B$-field respects the symmetries of 
$T^{1,1}$ and we will
assume that this is the proper $B$-field where the string is coupled
to. 
$h$, $K$, $f$ and $k$ are some functions of $\t$ whose form
can be found in \cite{KS}.
For our purpose, we record their asymptotic form
for large $\tau$,
\be
h = e^{-\frac{4 \t}{3}} (4 \t -1) + O(\t^2 e^{-\frac{10 \t}{3}} ) ,
\qquad
K =2^{1/3} e^{-\t/3}(1-  \frac{4 \t}{3} e^{-2 \t}) + O( e^{-\frac{2
    \t}{3}}),
\ee
\be f\rightarrow \frac{\tau -1 }{ 2} - \t e^{-\t} +O(\t e^{-2 \t}) \ , \qquad
k \rightarrow \frac{\tau -1 }{ 2} + \t e^{-\t} +O(\t e^{-2 \t}).\qquad\,\,\quad
\ee
In this limit, the metric becomes
\be
ds^2 = h^{-1/2}(r) dx^2 + h^{1/2}(r) ds_6^2,
\ee
where the radial variable is defined by
\be
r^3 = r_s^3 e^\t
\ee
for some resolved scale $r_s$. The warp factor is
\be
h = \frac{1}{r^4} \left(\log \frac{r}{r_s} -\frac{1}{4} \right) +
o\left( \frac{1}{r^{10}}  (\log \frac{r}{r_s})^2 \right)
\ee
and $ds_6^2$ is the cone metric  over
$T^ {1,1}$
\be
ds_6^2 = dr^2 + r^2 ds^2_{T^{1,1}}.
\ee
The $B$-field behaves
\be
B = O(\log \frac{r}{r_s}) (s_1 d \th_1 d \phi_1 -s_2 d \th_2 d \phi_2)  .
\ee
Putting $Y =1/r$, we
have  near the boundary $Y=0$
\bea
G_{\m\n} &=&  \frac{h_{\m\n}}{Y^2 \sqrt{\log Y}} \left(1+ O(\frac{1}{\log
    Y}) \right), \\
G_{ij} &=& k_{ij} \frac{\sqrt{\log Y}}{Y^2 } \left(1+ O(\frac{1}{\log
    Y}) \right),
\eea
and
\be
B_{ij} =O (\frac{\log Y}{Y^2}).
\ee
Here $h_{\m\n} =\eta_{\m\n}$ and $k_{ij}$ can
be worked out using the metric of $T^{1,1}$. These details will not be
important for us.
Note that the metric \eq{KS-metric} is of the form \eq{gen-metric} and
so it satisfies the condition \eq{GF}.

The Hamilton-Jacobi equation \eq{hj} is replaced by
\be \label{hj1}
(\log Y) k_{ij} J_1{}^\a \partial_{\a}Y^{i} J_1{}^\b \partial_{\b}Y^{j}
+  h_{\m\n} J_1{}^\a \partial_{\a}X^{\m} J_1{}^\b    \partial_{\b}X^{\n}
 = (\log Y)  k_{ij}\partial_1 Y^i \partial_1  Y^j+
 h_{\m\n} \partial_1  X^\m \partial_1  X^\n .
\ee
The string boundary condition is given by the same Dirichlet condition
\eq{dirichlet} and mixed boundary condition \eq{bci}. For a string
terminating on the boundary, we have $Y^i(\s_1, 0) =0$. To get rid of
the first term on the RHS of \eq{hj1}, we require that $\del_1 Y^i
(\s_1,0) = o (1/ \sqrt{\log Y})$. This also implies that the
$B$-term in the mixed boundary condition
\be
i B^k{}_l \del_1 Y^l = o(1).
\ee
The Hamilton-Jacobi equation in the limit $Y\to 0$ makes sense if
$J_1{}^\a \partial_{\a}Y^{i}(\s_1,0) $ is of the order of $1/ \sqrt{\log Y}$.
Therefore, we can drop the $B$-term in the mixed boundary
condition \eq{bci} and write
\be
J_1{}^\a \partial_{\a}Y^{i}(\s_1,0) = \frac{1}{\sqrt{\log Y}}
\L^i{}_j \dot{y}^j(\s_1).
\ee
The Hamilton-Jacobi equation finally gives
\be \label{loop-cond}
h_{\m\n} \dot{x}^\m \dot{x}^\n = k_{ij} \L^i{}_m \L^j{}_n \dot{y}^m
\dot{y}^n.
\ee

Now we examine the structure of UV divergences. For the area part,
it is easy to see that
we get the same linear divergence \eq{A-tilde} as before and so
$\tilde I_A$ is finite if the loop condition \eq{loop-cond} is satisfied. 
As for
the $B$-field, since
$\del_2 (B_{ij} \del_1 Y^i) Y^j $ is of the order of $\log Y/Y$,
therefore
\be
\tilde I_B  \sim (\log \e)^2.
\ee
This is a new divergence which can not be cancelled with the Legendre
transform.

\section{Discussions}

In this paper, we have  analysed of the structure of
UV divergences in the Wilson loop from the  supergravity point of
view by including the effect of a non-trivial metric and a NSNS
$B$-field.  We find that in general there can be new
divergences which  cannot be
cancelled with the Legendre transform.
We also find that when certain  conditions
are satisfied by the $B$-field and the metric,
the leading UV divergence becomes a linear one and this can be cancelled away
by choosing the boundary condition of the string
appropriately. In general there may still be divergences associated with the
$B$-field, and if they do exist, there is no way to cancel them with
the Legendre transform. This is similar to the
result of \cite{myers} which analysis the
effect of a nontrivial dilaton on the structure of UV divergences in
Wilson loop.
We conclude that Legendre transform is at best capable of
cancelling only linear UV divergences, but is incapable to cancelling any
subleading divergences which may be present, no matter whether it is
due to the dilaton or the NSNS $B$-field.

We have been concentrating on the structure
of UV divergences associated with the string minimal surface. For
Wilson loop in higher representations, a more suitable  dual
description  is in terms of D3-branes or D5-branes
\cite{df,yamaguchi,GP,OS,HP,GP2,kumar,dgrt}.
Presumably the correspondence will continue to hold for a more general
class of gauge/gravity duality. It will  be interesting to
analyze the structure of the  UV divergences there and to
derive the corresponding boundary conditions for  the corresponding
D-brane description.

Our analysis is performed on the supergravity side. It is an interesting
question to check and confirm the form of the loop constraint
\eq{constraint-final} from the field theory perspective. 
To do this, one need to know
the form of the Wilson loop operator that is dual to the supergravity
computation. In the simplest case where the field theory has the same
number of (adjoint) massless scalar with the dimension of the internal
manifold, the natural candidate for the operator is a direct
generalisation of \eq{wilsono}. However, the field theory may have       
different number of scalar fields in general. This is the    
case, for example, in the quiver theories that are dual to backgrounds with
Sasaki-Einstein spaces \cite{SE1,SE2}. There the form of the Wilson loop
operator is unknown. In this example one may try to exponentiate a product
of the bifundamental fields in order to construct the Wilson loop. But
since scalar field has dimension one in four dimensions, one needs to
compensate the dimension with another dimensional quantity. This is not  
completely clear what it might be in a conformal theory. It will be
interesting to analyze this further and to construct the Wilson loop
operator for these theories.

Finally we end with some remarks on the  form of the Wilson loop operator
in the 3-dimensional $\cN=6$ supersymmetric Chern-Simons theory \cite{ABJM}, where
recently the correspondence of Wilson loop has been analysed
\cite{BT,plefka,chen,rey2} (see also \cite{kluson} for related
discussions).
The ABJM theory has a $U(N)\times U(N)$ gauge and opposite levels $k$
and $-k$.  The matter fields are bifundamental scalar fields $A_1, A_2$ in the 
representation $(\bN, \bbN)$  and anti-bifundamental fields $B_1, B_2$
in the representation $(\bbN, \bN)$ and fermions. 
On the field theory side, a  Wilson loop operator which
couples to a certain bilinear combination of the bifundamental fields
has been considered
\be \label{wils-ex}
W[C] =\frac{1}{N}\Tr \,P\,
\exp\left[\oint_C d\t \left(iA_\mu \dot x^\mu+
\frac{2 \pi}{k} |\dot{x}| M_I{}^J Y^I Y^\dagger_J \right)\right], 
\ee 
where $Y^I =(A_1,A_2, \bar{B}_1, \bar{B}_2)$ and 
the curve $C$ is a straight line or a circle.
For the special
case where $C$ is spacelike and $M={\rm diag}(1,1,-1,-1)$, the operator is 1/6
BPS. In this case the UV divergences of 
this operator cancelled in the perturbation theory. It was also argued
\cite{plefka} that this 1/6 BPS Wilson loop operator describes
a string smeared over a $CP^1$ in $CP^3$. The smeared string perserves a $SU(2)
\times SU(2)$ subgroup of the $SU(4)$ isometry, which is precisely the
amount of R-symmetry preserved by the operator \eq{wils-ex} for this
particular choice of $M$. 
As a smeared configuration,
one would not expect to have a relation like \eq{bc-final} to relate
the worldsheet boundary conditions with the couplings of
the scalar fields in the Wilson loop. In general one may consider
localized string in $CP^3$ and ask how  it's boundary condition
appears in the Wilson loop. We will consider a natural proposal in the
following. However it turns out the correct operator has to be more
complicated than this.

To describe the string theory on $CP^3$ (see for example,
\cite{harmark}), it is convenient to use the complex coordinates $w^I$ 
\be
\sum_{I=1}^4 w^I \bar{w}^I =1,
\ee
subjected to the constraint
\be
\sum_{I=1}^4 (w^I \del_\a \bar{w}^I -\bar{w}^I\del_\a w^I )=0 , \quad
\a=1,2.
\ee
This construction is a realization of the Hopf fibration since 
the first constraint describes a $S^7$ and 
the second constraint describes a $U(1)$ symmetry which 
reduces the embedding to $CP^3$. 
Using this description, one can think about the transverse space to
the boundary spacetime $\bR^3$ as described by the four coordinates
$Z^I := Y w^I$ where $Y$ is the radial coordinate of $AdS_4$. In terms
of $Z^I$, we have $\sum_{I=1}^4 Z^I \bar{Z^I} = Y^2$ and
\be\label{Z2}
\sum_{I=1}^4 (Z^I \del_\a \bar{Z}^I -\bar{Z}^I\del_\a Z^I )=0 , \quad
\a=1,2.
\ee 
The string boundary
condition is then given by the three Dirichlet condition for the
longitudinal coordinates 
and the eight Neumann boundary conditions
\be \label{bc1}
J_1^\a \partial_\a Z^I(\tau,0)  = \dot{z}^I(\tau), \quad I=1, \cdots, 4 .
\ee
Note that the boundary condition \eq{bc1} is consistent with the
constraint in \eq{Z2} since $Z^I (\tau, 0) =0$. 
In terms of
real coordinates
$Z^1 =Y^1+ i Y^5, Z^2 =Y^2+ i Y^6, 
Z^3 =Y^3+ i Y^7,  Z^4 =Y^4+ i Y^8$, 
the embedding reads $\sum_{i=1}^8 (Y^i)^2 =Y^2$ and
\be \label{Y1}
\sum_{I=1}^4 (Y^I \del_\a Y^{I+4} -Y^{I+4} \del_\a Y^I )=0. 
\ee
The boundary condition reads
\be 
J_1^\a \partial_\a Y^i(\tau,0)  = \dot{y}^i, \quad i=1,\cdots, 8,
\ee
where
$z^1 =y^1+ i y^5,  z^2 =y^2+ i y^6, 
z^3 =y^3+ i y^7, z^4 =y^4+ i y^8. $
 
To write down the Wilson loop, we note that due to the presence of 
the product gauge group,  there are 
two independent Wilson loops one can write down.  Let us concentrate
for the moment on
the first $U(N)$, one can form adjoint fields by multiplying the
bi-fundamental fields in a certain order.  It is natural to consider 
\be \label{proposal}
W = \frac{1}{N} \Tr P \exp \left(
\oint_C d\tau (i A_\m \dot{x}^\m + \dot{a}_{a b} A_a \bar{A}_b   
+  \dot{b}_{a b} \bar{B}_a B_b)\right) 
\ee
where $C$ is a general spacelike curve. This operator is invariant
under arbitrary reparametrization $\tau \to \tilde{\tau}$, including
orientation reversing ones.  
Since scalar fields in three-dimensions is of dimension half, the variables
$a^{a b}$ and $b^{a b}$ are of length dimension and therefore it
make sense to try to identify them with the  boundary variables $z^I$
in \eq{bc1}. Since $A_a$ (or $B_a$) is a doublet of $SU(2)_1$, $A_a \bA_b$
(or  $\bar{B}_a B_b$) 
contains a singlet and a triplet of $SU(2)_1$. Our proposal is to identify
\be \label{iden1}
\dot{a}_{a b} = \frac{2 \sqrt{2} \pi}{k}   \sum_{i=1}^4 (\s^i)_{a b} \dot{y}^i,
\quad
\dot{b}_{a b} = \frac{2\sqrt{2}  \pi}{k} \sum_{i=1}^4 
(\s^i)_{a b} \dot{y}^{i+4}
\ee 
where $\s^i= (\tau^1,  \tau^2,  \tau^3, 1)$ and $\tau^{1,2,3}$ are the
Pauli matrices. Note that the ABJM theory is manifestly invariant under
$SU(2)\times SU(2)$ of the $SU(4)$ R-symmetry. Therefore 
\eq{proposal} respects this symmetry if we assign 
$(y^1,y^2, y^3)$ (respectively $(y^5,y^6, y^7)$) 
to be a triplet and $y^4$ (respectively $y^8$) to be a singlet
$SU(2)_1$ (respectively  $SU(2)_2$). 
For convenience, 
we have put a factor of $2\sqrt{2} \pi/k$ above since
the propagator of the gauge bosons and the scalar field is
different. This turns out to be a convenient normalization 
in perturbation theory. 
We remark  that the identification \eq{iden1} can also be written as
\be \label{iden2}
\dot{a}_{a b} + i \dot{b}_{a b} =  
\frac{2 \sqrt{2}\pi}{k}   \sum_{I=1}^4 (\s^I)_{a b} \dot{z}^I
\ee
and our proposal for the Wilson loop operator that is dual to a string
with the boundary condition \eq{bc1} is
\be \label{wilson-prop}
W = \frac{1}{N} \Tr P \exp \left[
\oint_C d\tau \Big(i A_\m \dot{x}^\m + \frac{2 \pi}{k} 
\sum_{I=1}^4 \dot{z}^I \bar{R}^I     +  \dot{\bz}^I R^I  
\Big)\right].
\ee
Here $R^I$ is the composite scalar 
$R^I := (\cA^I + i \cB^I)/\sqrt{2}$
where $\cA^I := A_a (\s^I)_{ab} \bA_b$, $\cB^I:= \bB_a (\s^I)_{ab}
B_b$ .

By doing a perturbative computation as in, e.g. \cite{plefka,chen,rey2}, one can
show that  the Wilson loop is in general linear divergent:
\bea
\sim \frac{N^2}{k^2   \e}  
\int d\t_1   ( \dot{x}(\tau_1)^2- \dot{y}(\tau_1)^2).
\eea
Therefore the divergence cancels if the loop constraint
\be
 \dot{x}^2= \dot{y}^2
\ee
is satisfied. The fact that we obtain precisely the same constraint as 
obtained from the Hamilton-Jacobi analysis 
provides some support that
the ansatz \eq{wilson-prop} correctly encodes the boundary conditions of the dual
open string. However this cannot be correct due to a mismatch. In
fact, a half BPS string configuration 
which is localized at a point in $CP^3$  has been considered in
\cite{plefka,chen,rey2}. 
One can show that there is no choice of $\dot{z}^I$ to make 
\eq{wilson-prop} half BPS. Even worse, it is easy to 
show, for the ansatz \eq{wils-ex} which is coupled to a bilinear of
scalars,  there is no choice of the Hermitian matrix $M$ so
that there is 1/2 unbroken supersymmetry. Therefore the correct Wilson loop
operator that is dual to localized string must be more complicated.
The understanding of this will be very interesting.

\section*{Acknowledgements}

We would like to thank  Jaume Gomis, Asad Naqvi and
Douglas Smith for stimulating discussions and useful comments.
The research of CSC is  partially supported by EPSRC and STFC.
The work of DG is supported by an EPSRC studentship.

\end{document}